# Adaptive Accessible AR/VR Systems

PRADIPTA BISWAS [1]*

Indian Institute of Science, Bangalore, India

PILAR ORERO

University of Barcelona, Spain

MANOHAR SWAMINATHAN

Microsoft Research, Bangalore, India

KAVITA KRISHNASWAMY

University of Maryland, Baltimore Country, USA

PETER ROBINSON

University of Cambridge, United Kingdom

Augmented, virtual and mixed reality technologies offer new ways of interacting with digital media. However, such technologies are not well explored for people with different ranges of abilities beyond a few specific navigation and gaming applications. While new standardization activities are investigating accessibility issues with existing AR/VR systems, commercial systems are still confined to specialized hardware and software limiting their widespread adoption among people with disabilities as well as seniors. This proposal takes a novel approach by exploring the application of user model-based personalization for AR/VR systems to improve accessibility. The workshop will be organized by experienced researchers in the field of human computer interaction, robotics control, assistive technology, and AR/VR systems, and will consist of peer reviewed papers and hands-on demonstrations. Keynote speeches and demonstrations will cover latest accessibility research at Microsoft, Google, Verizon and leading universities.

## ACM CLASSIFICATION KEYWORDS

H.1.2 User/Machine Systems: Human factor

---

# 1 BACKGROUND

Research on mainstream interactive devices and accessible computing are often disjoint. Accessible computing or assistive technology is considered a small segment of the market, and mainstream products and services are rarely investigated for a wider range of abilities than typical 'average' users. An international survey organized by the World Health Organization (WHO) and the World Bank reports that more than 1 billion people, about 15% of the world's population, have some form of disability and by 2020 there will be 2 billion people aged 60 or older [33]. According to WHO, more than 1 billion people need assistive products today and more than 2 billion individuals are expected to need at least one assistive product by 2030 globally, but only one in 10 have access to them [33]. Recent advances in artificial intelligence, interactive systems and graphics processing units mean that we can now use AR and VR technologies in smartphones and can download software code to train complicated convolutional neural networks for object detection including analysis of facial expressions. However, there has been little reported work on the use of Augmented, Virtual and Mixed Reality (AR, VR and MR) technologies by users with different ranges of abilities, except for a few gaming, navigation and rehabilitation applications. Inclusive AR/VR systems have tremendous potential for changing the lives of people with different ranges of abilities in providing convenient facilities for communication, education, and rehabilitation.

This workshop will explore novel approaches to bringing the latest developments in computing technologies to users who often miss out advantages in information technology due to their limited range of abilities and are excluded from mainstream society. In particular, the workshop plans to bring together developers of AR/VR and assistive technologies and explore new applications and services of AR/VR media for users with different ranges of abilities.

## 1.1 Existing Work on Accessible AR/VR systems

Researchers already explored use of immersive media for people with different range of abilities including physical, hearing, learning and visual impaired users. Pose estimation [32], semantic segmentation [31], and depth estimation [16] are helpful augmented reality features that can help to improve access to assistive technology for everyone with digital information overlaid in parallel with the real world camera as an image, audio, or video for impactful and creative application developments. Virtual reality technology is reported to be useful for teaching cognitive spatial map to young users with physical impairment [25], who cannot move around themselves. It is used to teach spatial navigation to wheelchair users. Virtual and augmented reality technologies are also used for training physical movements to young users [20] as well as older patients as part of their stroke rehabilitation program [28, 19]. Burdea [3] classified the VR rehabilitation projects into orthopedic, imitation training and cognitive rehabilitation projects. Rutgers Ankle [12] is an example of orthopedic rehabilitation where patients were trained to fly an aircraft in VR through ankle movement with haptic feedback and a real time graph showed the range of motion and torque of ankle movement. Holden [15] used a VR hand movement simulator to help stroke patients exercising hand movements. A VR flight simulator [14] was used for patients afraid of flying while a VR classroom [26] was used to detect and treat attention deficit and hyperactivity disorder (ADHD) in children.

## 1.2 Multimodal Interaction

One of the earliest examples of immersive media, the Sensorirama simulator provided haptic, auditory and tactile feedback in terms of gust of wind for simulating a bike-riding experience in New York. In a recent paper, Murthy [23] reported various multimodal interaction techniques available for AR/VR media. For users with different range of abilities, multimodal feedback can include a wide range of users compared to traditional AR/VR with only visual output. The use of spatial audio interaction tools and techniques to include people with a range of vision impairments in mainstream video games has been explored and a plugin is available at the Unity store to enable game developers to build inclusion from the ground up [29]. SeeingVR is a recent research toolkit [35] for making VR accessible to people with low vision. Liarokapis [21] developed an augmented virtual museum tour while Richard [25] presented an AR based educational tool for cognitively disabled students with auditory and olfactory feedback. Lin [22] presented a study involving three children with motor impairment and an augmented reality game to motivate them towards physical activity. The study generated promising results and Lin [22] further reported that the flexible AR system is adopted in elementary and junior high schools for students with paralysis and cerebral palsy. For users with severe motor impairment, Sharma [27] reported a set of user studies with an eye gaze controlled video see through system integrated with a robotic manipulator (figure 1) for pick and drop tasks. Gerling [11] undertook detailed qualitative study on usability of VR headset for wheelchair users and proposed to use head and wheelchair movement and punching action with VR environment. They had to automate and adapt interaction with a VR gaming environment when users found moving head or wheelchair difficult and emphasized on personalizing interaction based on range of abilities of users.

Several EU projects are investigating inclusive and interactive AR/VR systems. The EU ImAc project [6] investigated accessibility of immersive media and developing accessible closed captioning system for VR media, sign language and also audio description. The Aurora project [1] is investigating use of robotics for enhancing social inclusion for children with autism. The EU Traction [9] project is investigating the use of VR for social cohesion taking opera as the cultural context. The EU SoClose [8] project is investigating the use of VR with migrants and asylum seekers, as a way to co-create and share experiences. In the research collaboration between the University of Maryland, Baltimore County (UMBC) and the University of Washington, a suite of five different accessible robotic web-interfaces were developed with augmented reality and multimodal interaction, multi-view perspective, and machine learning algorithms to model and observe predictive user intentions to assist people with disabilities [18]

## 1.3 Accessibility Standards and Guidelines for AR/VR Technologies

The XR Access [34] community investigated accessibility issues of immersive media and as of August 2020, identified 19 different areas to improve and promote accessibility of immersive media. However, the XRAccess community mainly looked at immersive media but not augmented reality in detail, which is often easier and affordable to deploy in large scale than virtual reality systems. The ITU Study Group 6, 9 and 16 investigated accessibility landscape of subtitles, cable and broadband TV and multimedia systems respectively and formed a joint Inter Rapporteur Group on to consolidate effort on developing accessible audio-visual media. ISO JTC1 SG35 WG6 is developing a standard ISO 20071-30 for accessibility in immersive environments. In order to accommodate the ever-growing user base of Oculus products, during mid-November 2020, Facebook issued a set of guidelines on UX/UI, control, interaction, caption, subtitle and so on for designing accessible VR applications. Facebook also initiated a research group on bystander security for developing inclusive AR applications.

## 1.4 Personalization

While many existing AR/VR related research so far looked at specific applications (like stroke rehabilitation) for people with different range of abilities, researchers on accessible computing already took effort to personalize interaction mainly for traditional audio visual media. This workshop will investigate interaction issues of immersive media in details with respect to a wide range of social, physical and cognitive abilities as well as prospect of integrating existing research on interface personalization with Augmented, Virtual and Mixed Reality systems. This workshop aims to investigate integrating intelligent predictive algorithms with immersive media through user modelling framework.

It may be noted that an impairment does not always create a specific disability – for example, a person with hearing impairment may also need bigger font size and enhanced colour contrast to read a subtitle and a young adult with cerebral palsy require less number of elements on a screen due to limited cognitive ability. Accessibility systems and services often cater a wider range of audience than its speculated end users for example audio books for visually impaired are often used by drivers and subtitles are useful for foreign language speakers, or the general public consuming media in public spaces. Quality and usability of an assistive technology itself is also important for its market acceptance as like other technologies – an unsynchronized subtitle or a virtual keyboard with small key size often impede its use by persons with disability. Considering these facts, we propose to offer accessibility services as a personalization service for people with different range of abilities. A user profile can store intended settings of a user like font size, volume, and so on and different services including accessibility ones can be personalized based on the user profile. Interface personalization is mainly explored in the domain of content personalization and developing intelligent information filtering or recommendation systems based on user profiles. In most of those systems content (or information) is represented in a graph like structure (e.g. ontology or semantic network) and filtering or recommendation is generated by storing and analyzing users' interaction patterns. Existing research also investigated personalizing user interfaces beyond content personalization for users with different range of abilities. A few representative and significant projects on interface personalization are the SUPPLE project at University of Washington [10], and AVANTI project [30] and the EASYTV [5] for people with disabilities. The SUPPLE project personalizes interfaces mainly by changing layout and font size for people with visual and motor impairment and also for ubiquitous devices. However, the user models do not consider visual and motor impairment in detail and thus work for only loss of visual acuity and a few types of motor impairment. The AVANTI project provides a multimedia web browser for people with light, or severe motor disabilities, and blind people. The EASYTV project personalized the interaction with media content in all web-based devices. The Global Public Inclusive Infrastructure [13] in USA and its EU counterparts like Cloud4All [4] and Prosperity4All [7] projects proposed an inclusive infrastructure for developing, publishing and searching appropriate inclusive systems for people with different range of abilities. These projects did a commendable job in specifying the scopes and objectives of a large scale inclusive infrastructure but implementation-wise the projects are still at preliminary stage and only developed a few technology demonstrators, which are yet to leverage the whole framework. More specifically, these projects do not have any objective mapping between users' range of abilities and interface parameters and works mainly based on users' explicitly stated preferences. The GPII auto-personalization feature is a step towards this mapping but yet to be explored for multiple platforms, in particular AR/VR systems.

### 1.5  Common User Profile

User models provide a solution to adapt existing user interfaces for a wide range of users. A user model can be defined as a machine-readable representation of a user while an instantiation of a user model is termed as a user profile. The Inclusive User Model [2] at University of Cambridge simulated interaction in terms of visual perception, cursor movement and so on for users with visual, auditory and motor impairment. The simulation was later used for both automatic accessibility evaluation and personalization of user interfaces. The EU VUMS (Virtual User Modelling and Simulation) [17] was an ensemble of four projects that explored developing a common user profile format for personalizing both electronic and physical user interfaces for people with different range of abilities. In the context of personalization and accessibility, a common user profile format may have following advantages

1. Personalizing user interface content and layout for different applications after creation of a single user profile
2. Offering accessibility services to all devices and platforms after creation of a single user profile
3. Sharing personalized content and interface across different platform and devices to improve usability
4. Adapting quality of accessibility services (e.g. font size of subtitle) across multiple media
5. Sharing personalization metadata among service produces like website or content developers

However, sharing information about user always possess security risk and unintended use not authorized by end users. Implementation of the common user profile should take care of security aspect and local regulation and legislation. If the format and details of common user profile is agreed and shared among service providers, then sharing of actual content may not be necessary, rather the personalization algorithms can run on user profile stored on local machines. Standardization and sharing of only the format definition across service providers will enable personalization without taking the risk of sharing details of individual user.

## 2  MOTIVATION

This workshop aims to extend existing research on user modelling and personalization for AR/VR systems enhancing usability of such systems for users with different range of abilities. A similar workshop was organized in 2012 at ACM Intelligent User Interface conference on enhancing accessibility of digital content and led to two editions of a Springer book and standardization initiatives at International Telecommunication Union in the form of focus groups on accessibility, smart TV and general multimedia content. In recent time, the organizers are involved with various accessibility as well as AR/VR related projects funded by European commission, major software and telecommunication industries and defense agencies. Earlier research already explored use of AR/VR technologies for navigation and rehabilitation. In the present Covid 19 pandemic situation, online training and electronic learning platforms turned too important. Teaching and learning of science can take a whole new dimension with AR /VR technologies. Immersive learning has proven to be very beneficial in the case of learning new languages, an area of great difficulty for children with learning disabilities. Gamification and interactivity will improve student engagement and learning outcome. While VR and full immersion still require high end computing stations in the form of high-end smartphone, wearable headsets and high end graphics processing units, low-cost optical and video see through interfaces are easier to scale up to large population. However, like any other technology, large scale adoption of AR/VR technology requires it to be intuitive, easy to use and maintain and adding new values compared to traditional interactive systems. This workshop will aim to bring together AR/VR developers and accessibility experts together and explore innovative ways to produce immersive inclusive content for education and ICT applications. Figure 1 below shows an eye gaze-controlled Video See Through (VST) Human Robot Interaction (HRI) system for users with severe speech and motor impairment while figure 2 shows subtitles for a VR system.

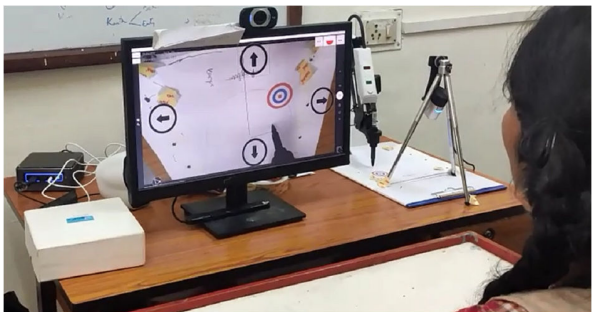

**Figure 1.** Eye gaze controlled HRI using AR for users with severe speech and motor impairment

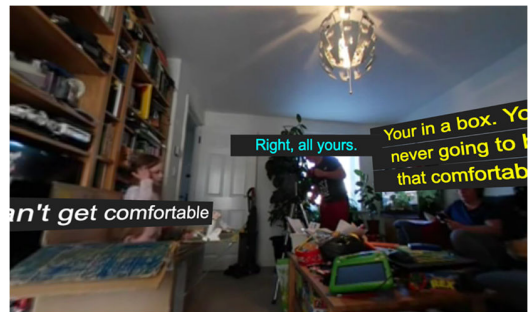

**Figure 2.** Subtitle in immersive environment for users with hearing impairment

Systems and services developed for elderly or disabled people often finds useful applications for their able bodied counterparts – a few examples are mobile amplification control, which was originally developed for people with hearing problem but helpful in noisy environment, audio cassette version of books originally developed for blind people, standard of subtitling in television for deaf users and so on. In this specific context, the eye gaze controlled see through display (figure 1) is also used as an interactive Head Up Display in cars [24] and the subtitles in figure 2 found applications in immersive learning system for able-bodied users as well (figure 3).

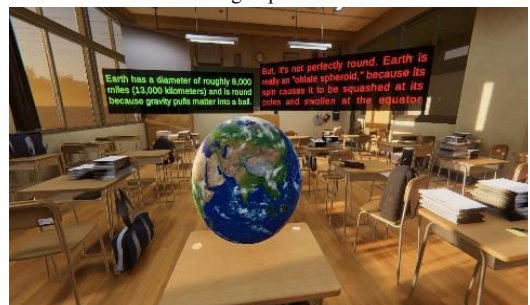

**Figure 3.** Immersive eLearning system with subtitles

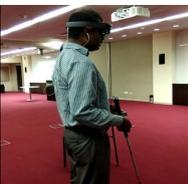

**Figure 4.** Spatial Audio

Immersive visual environments can also be made accessible by the use of spatial audio technology that is now available in commodity smartphones. The use of a spatial audio cloud surrounding the user along with a set of interaction tools that extract and present the relevant information in the environment through spatial audio can make VR and AR environments accessible to people with vision impairments. A toolkit that enables designers to build in such features is available as a Unity plugin [29]. A person with vision impairments using a HoloLens to explore the mixed reality environment through spatial audio is presented in the Figure 4.

## 3 ORGANIZERS

**Pradipta Biswas** is an assistant professor at the Centre for Product Design and Manufacturing (CPDM) and associate faculty at the Robert Bosch Centre for Cyber Physical Systems (RBCCPS) of Indian Institute of Science. His research focuses on user modelling and multimodal human-machine interaction for aviation, automotive and assistive technology. Pradipta is a Co-Chair of the IRG AVA and Focus Group on Smart TV at International Telecommunication Union. He is a member of the UKRI International Development Peer Review College, Society of Flight Test Engineers and was a professional member of the British Computer Society, Associate Fellow at the UK Higher Education Academy and Royal Society of Medicine. Earlier, he was a Senior Research Associate at Engineering Department, Research Fellow at Wolfson College and Research Associate at Trinity Hall of University of Cambridge. He completed PhD in Computer Science at the Rainbow Group of University of Cambridge Computer Laboratory and Trinity College in 2010 and was awarded a Gates-Cambridge Scholarship in 2006.

**Pilar Orero (PhD, UMIST)** works at the Universitat Autònoma de Barcelona (Spain), in the TransMedia Catalonia Research Group. She is a world-leading scholar in media accessibility with vast experience in standardization and policy-making, she is a scientific/organizing committee member of many conferences, including Media4All, ARSAD, and Video Games for All. She has delivered, upon invitation, more than 15 plenary lectures and 30 guest lectures all over the world, including at the 9th United Nations Conference of the States Parties to the Convention on the Rights of Persons with Disabilities (New York, 2016). She either coordinated or participated in more than 40 national and international research projects, of which more than 20 were on media accessibility.

**Manohar Swaminathan** (also known as Swami Manohar) is a Principal researcher at Microsoft Research India, where he is part of the Technologies for Emerging Markets group. Manohar is an academic-turned technology entrepreneur-turned researcher with a driving passion to build and deploy technology for positive social impact. He has a PhD in CS from Brown University, was a Professor at the Indian Institute of Science, and has co-founded, managed, advised, and angel-funded several technology startups in India. He has guided over 40 graduate students and has more than 40 refereed publications. His current research interests are broadly in two areas: robust sensing for IoT applications in agriculture and in pollution monitoring and ludic design for accessibility. The latter covers a broad set of interdisciplinary research topics, including computational thinking for children who are blind, video games for the vision impaired, and gaze-tracked interfaces for children with sensory motor impairments.

**Kavita Krishnaswamy** is a Ph.D. candidate advised by Dr. Tim Oates at the University of Maryland, Baltimore County (UMBC). She received her Bachelor's degree in Mathematics and Computer Science at UMBC while maintaining a perfect 4.0 grade-point average. She lives with a progressive neuromuscular disorder called spinal muscular atrophy (SMA) and is only able to move the index finger of her right hand. As a roboticist, her work focuses on the development of robotic systems to increase independence for people with disabilities with accessible interfaces with augmented reality. Along with being named among the "25 Women to Watch" by the Baltimore Sun and "5 Women Under 40 to Watch" by New Mobility magazine, Kavita has been recognized as a Google Lime Scholar, Microsoft Research Fellow, Ford Foundation Predoctoral Fellow and National Science Foundation Graduate Research Fellow.

**Peter Robinson** is Professor of Computer Technology in the Department of Computer Science and Technology at the University of Cambridge, where he is part of the Rainbow Research Group working on computer graphics and interaction. Professor Robinson's research concerns new technologies to enhance communication between computers and their users, and new applications to exploit these technologies. For many years he led work on video and paper as part of the user interface using video projection and digital cameras. More recently he has worked on affective computing, the inference of people's mental states from social signals and the expression of emotions by robots and cartoon avatars.

## 4 LINK TO WEBSITE

The website describes the aim of the workshop and identifies a set of research questions. It has links to a one-page flyer, organizers' and keynote speakers' homepages. A demonstration video is embedded in the website and also submitted as supplementary material. The website is available at https://cambum.net/Inclusive_ARVR/ The website itself is evaluated with *A-Checker by Inclusive Design* WCAG validation tool.

## 5 PRE-WORKSHOP PLANS

The workshop website will be published through organizers' host institutes (Indian Institute of Science, Universitat Autònoma de Barcelona, University of Maryland, Microsoft Research and University of Cambridge) broadcast mailing list. The workshop will also be publicized through International Telecommunication Union's accessibility related focus and study groups, International Standardization Organizations usability related study committees, ACM mailing lists and organizers' personal and professional social networking websites. It will be highlighted in the LEAD ME COST Action 19142 specialized on Media Accessibility.

## 6 WORKSHOP STRUCTURE

The workshop will have three keynote speeches, regular paper sessions and hands-on experience sessions where attendees can participate in latest interactive and inclusive immersive media development. An opening and closing plenary session will not only highlight research challenges but also touch upon issues with standardization, large scale development and deployment of inclusive immersive technologies. In particular, we aim for at least 10 paper presentations, each presenter will be given 10 minutes presentation slot. There will be four sessions each with one-hour duration with breaks for informal discussion in-between sessions. The closed captioning feature of MS Teams will be kept on. During informal sessions, participants can either create virtual rooms or join an existing virtual room for experiencing hands-on demonstration.

**Keynote speeches:** The workshop will have 3 keynote speeches, each having 30 minutes session. Mr Larry Goldberg, Head of Accessibility, **Verizon** media will present on "Virtual Reality meets Accessibility" while Sonali Rai, Senior manager **RNIB** (UK) will present on "Immersive Accessibility: what's next?". Mr Chris Patnoe, senior project manager at **Google**, will present on latest development on accessibility features for immersive media.

**Demonstration session:** A set of virtual machines with pre-loaded software will be set up before the workshop and the log in information will be shared through the website as well as registered email list. Attendees can experience eye gaze controlled VR sensor dashboard that they can operate without moving hands and an eye gaze controlled video see through display that can be used to operate cyber physical systems like robot and remotely operated ground and aerial vehicles. A demonstration will be set up of enforcing COVID 19 related social distancing in a VR digital twin of office spaces developed for a funded project by **British Telecom (BT)**. Interactive VR demonstrations will be exported using WebGL like the following web page (https://tinyurl.com/y2rhh2da)and linked to conference website. Colleagues from **Microsoft** will demonstrate the Unity plugin for spatial audio and accessible navigation system for blind using Microsoft Hololens. Participants will have the opportunity to create and develop a webpage that uses augmented reality directly in the web browser via TensorFlow.js from their own live webcam stream in real time. Specifically, participants will be introduced to the detection of real-time human pose estimation using the PoseNet model as well as the FaceMesh and HandPose models for tracking key landmarks on faces and hands, respectively. Using these three models, participants can create an augmented reality overlay for their live webcam stream in the webpage that can be saved as an image or video for sharing. Attendees can experience novel accessibility services for immersive media like subtitles with head movement tracking and various research challenges in terms of tracking, calibration, and interaction with AR/VR systems. In summary, we shall have demonstrations on

1. Eye gaze controlled AR/VR system for education, communication and rehabilitation of users with disabilities
2. AR system for human robot interaction for users with motor impairment
3. Eye gaze and head movement controlled subtitles for VR environment for users with hearing impairment
4. Spatial Audio in VR environment for users with visual impairment
5. Personalization of content of using an online common user profile format for users with disabilities.

A demonstration video is available at https://youtu.be/sP7MtnwdVlY The demonstration video shows eye gaze controlled video see through interface for human robot interface for users with severe speech and motor impairment, VR subtitle for an eLearning project, spatial audio for an accessible VR game and a VR digital twin with real time interactive sensor dashboard. The schedule of the workshop is furnished in Table 1 below. There will be 20 minutes break between each pairs of sessions.

Table 1. Schedule of the Workshop

| Session 1 (1 hr) | Session 3 (1 hr) |
|---|---|
| Opening Plenary, Introductions (10 mins) | 3 Paper Presentations (30 mins) |
| Keynote Speech (30 mins) | Keynote Speech (30 mins) |
| 2 Paper Presentations (20 mins) | |
| **Session 2 (1 hr)** | **Session 4 (1 hr)** |
| 3 Paper Presentations (30 mins) | 2 Paper Presentations (20 mins) |
| Keynote Speech (30 mins) | Concluding Remarks, Post Workshop Plan (40 mins) |

## 7 POST-WORKSHOP PLANS

Accepted papers will either be published as a special issue of a Scopus indexed open access journal. Authors will also be given an option to refrain from publication and keep copyright of their papers. A white paper from the workshop will be presented to various standardization organizations like ITU and ISO for considering to influence future international standards. With permissions from authors, video recordings of all presentations and hands on demonstrations sessions will be published in the website. The website will be linked to organizers' institutional websites like the Universitat Autònoma Digital Repository DDD and will be indexed in the Media Accessibility Platform MAP. The video materials will be made accessible by adding subtitles and audio captions. Using ACM Distinguished Speakers scheme, DST DUO Fellowship from India and industrial funded projects, all organizers will visit each other and take the work forward in terms of both developing new accessible AR/VR systems as well as influence future standards on AR/VR product development.

## 8 CALL FOR PARTICIPATION

In recent time, both artificial intelligent and interactive systems made tremendous progress. The workshop on Adaptive Accessible AR/VR Systems is planning to take a novel approach to bring these latest developments in computing technologies for users, who often miss out advantages in information technology due to their limited range of abilities.

**Format and goals of the workshop:** The workshop will have oral paper presentation and hands on demonstration session. The goals of the workshop are
1. Investigating accessibility issues with AR/VR systems
2. Discussing state-of-the-art in inclusive AR/VR systems
3. Exploring future prospect of mass deployment of AR/VR systems for people with disabilities

**Selection criteria:** Papers will be selected through double-blind peer review process with each paper being reviewed by at least two reviewers and one member of the organizing committee

**Requirements for position papers:** This workshop is seeking contributions in the form of regular research papers, case study reports, description of use cases for following topics
- Interactive AR/VR applications for users with different range of abilities
- AR/VR games for education and rehabilitation
- E-Learning or other ICT systems with AR/VR component

Papers should be between 5000 and 8000 words in length in ACM paper format.

At least one author of each accepted position paper must attend the workshop and all participants must register for both the workshop and for at least one day of the conference. Details of paper submission link and deadlines can be found at the conference website http://cambum.net/Inclusive_ARVR/

# REFERENCES

[1] Aurora Project, http://aurora.herts.ac.uk/, Accessed on 12/3/2020
[2] Biswas P. and Robinson P. and Langdon P. (2013), Evaluating Interface Layout for Visually and Mobility Impaired Users through Simulation, Universal Access in the Information Society (UAIS), 12 (1)
[3] Burdea GC and Coiffet P(2016), Virtual Reality Technology, Wiley 2016
[4] EU Cloud4All 2016, http://www.cloud4all.info/, Accessed on 17/4/2020
[5] EU EasyTV project (2020), https://easytvproject.eu/, Accessed on 09/09/2020
[6] EU ImAc project (2020), https://www.imac-project.eu/, Accessed on 09/09/2020
[7] EU Prosperity4All 2016, http://www.prosperity4all.eu/, Accessed on 17/4/2020
[8] EU SoClose project (2020), https://so-close.eu/, Accessed on 09/09/2020
[9] EU Traction project (2020), https://www.traction-project.eu/, Accessed on 09/09/2020
[10] Gajos K. Z., Wobbrock J. O. and Weld D. S. Automatically generating user interfaces adapted to users' motor and vision capabilities. ACM symposium on User interface software and technology 2007. 231-240.
[11] Gerling K., Dickinson P., Hicks K., Mason L., Simeone A. L., and Spiel K. (2020). Virtual Reality Games for People Using Wheelchairs. In Proceedings of the 2020 CHI Conference on Human Factors in Computing Systems (CHI '20).
[12] Girone, M., G.Buurdea , and M. Bouzit, 1999,  The 'Rutgers Ankle' Orthopedic Rehabilation Interface, in Proceedings of the ASME Haptics Symposium , DSC- Vol. 67 , ASME, 305-312.
[13] GPII 2016, http://gpii.net/, Accessed on 17/4/2020
[14] Hodges, L., B. Rothbaum, B. Watson, D. Kessler, and D. Opdyke, 1996, A Virtual Airplane for Fear of Flying Therapy, in Proc. IEEE VRAIS, 86-93
[15] Holden, M., and E. Todorov, 2002, Use of virtual Environments in Motor Learning and Rehabilation , in K. Stanney (Ed), The Handbook of Virtual Environments Technology, Erlbaum, Mahwah, NJ, 999-1026
[16] Jones, J. A. (2020). Egocentric depth perception in optical see-through augmented reality (Doctoral dissertation, Mississippi State University).
[17] Kaklanis N., Biswas P., Mohamad Y., Gonzalez M. F., Peissner M., Langdon P., Tzovaras D. and Jung C., Towards Standardization of User Models for Simulation and Adaptation Purposes, Universal Access in the Information Society (UAIS) 15(1)
[18] Krishnaswamy, K., Adamson, T., Cakmak, M., & Oates, T. (2018). Multi-perspective, multimodal, and machine learning for accessible robotic web interface. In proceedings of RESNA '18: Rehabilitation Engineering and Assistive Technology Society of North America. Retrieved from https://www.resna.org/sites/default/files/conference/2018/outcomes/Aghababa.html
[19] Kuhlen T. W. and Dohle C. (1995), Virtual reality for physically disabled people, Computers in Biology and Medicine, April 1995
[20] Lányi C. S., Geiszt Z., Károlyi P., Tilinger A. and Magyar V. (2006), Virtual Reality in special needs early education, International Journal of Virtual Reality, 5(3):1-10
[21] Liarokapis F., Sylaiou S., Basu A., Mourkoussis N. (2004), White M. and Lister P. F. (2004), An Interactive Visualisation Interface for Virtual Museums, The Eurographics Association 2004
[22] Lin C and Chang Y (2015), Interactive augmented reality using Scratch 2.0 to improve physical activities for children with developmental disabilities, Research in Developmental Disabilities 37:1–8
[23] Murthy lrd (2020), Multimodal Interaction for Real and Virtual Environments, ACM International Conference on Intelligent User Interfaces (IUI 2020)
[24] Prabhakar G., Ramakrishnan A., Murthy LRD, Sharma V. K., Madan M., Deshmukh S. and Biswas P. (2020), Interactive Gaze & Finger controlled HUD for Cars, Journal on Multimodal User Interface, Springer, 2020
[25] Richard E., Billaudeau V., Richard P., and Gaudin G. (2007), Augmented Reality for Rehabilitation of Cognitive Disabled Children: A Preliminary Study, IEEE Virtual Rehabilitation 2007
[26] Rizzo A. et al (2000), The Virtual Classroom: A virtual reality environment for the assessment and rehabilitation of attention deficits, CyberPsychology and Behaviour 3(3), 483-499
[27] Sharma VK, Saluja KPS, Mollyn V and Biswas P., Eye Gaze Controlled Robotic Arm for Persons with Severe Speech and Motor Impairment, ACM International Conference on Eye Tracking Research and Applications (ETRA 2020)
[28] Sisto S. A., Forrest G. F. and Glendinning D. (2002), VR Applications for Motor Rehabilitation After Stroke, Topics in Stroke Rehabilitation 8(4)
[29] Spatial 2019: https://assetstore.unity.com/packages/templates/systems/responsive-spatial-audio-for-immersive-gaming-a-microsoft-garage-144702
[30] Stephanidis C. and Constantinou P. Designing Human Computer Interfaces for Quadriplegic People. ACM Transactions On Computer-Human Interaction 10.2 (2003): 87-118.
[31] TensorFlow. 2020. BodyPix. https://github.com/tensorflow/tfjs-models/tree/master/body-pixWHO, 2011. Global Health and Ageing, Available at: <https://www.who.int/ageing/publications/global_health.pdf, Accessed 27 September 2020
[32] Tripathi, S., Ranade, S., Tyagi, A., & Agrawal, A. (2020). PoseNet3D: Unsupervised 3D Human Shape and Pose Estimation. arXiv preprint arXiv:2003.03473.
[33] WHO, 2016. Priority Assistive Products List. Available at: https://apps.who.int/iris/bitstream/handle/10665/207694/WHO_EMP_PHI_2016.01_eng.pdf, Accessed 27 September 2020
[34] XRAccess Initiative, https://xraccess.org/, Accessed on 10/09/2020
[35] Zhao 2019:  Yuhang Zhao Ed Cutrell Christian Holz Meredith Ringel Morris Eyal Ofek Andrew D. Wilson, SeeingVR: A Set of Tools to Make Virtual Reality More Accessible to People with Low Vision, CHI 2019 | May 2019, ACM